\documentclass[thmsa,onecolumn,ukenglish,12pt,a4paper,oneside,final,notitlepage]{article}
\usepackage[centertags,sumlimits,nointlimits,namelimits,reqno]{amsmath}
\usepackage{sw20lart}

\input{tcilatex}
\begin{document}

\title{Ion-Acoustic Waves in Unmagnitized Collisionless Weakly Relativistic
Plasma using Time-Fractional KdV Equation}
\date{}
\author{El-Said A. El-Wakil, Essam M. Abulwafa, \and Emad K. El-shewy and
Abeer A. Mahmoud \\
Theoretical Physics Research Group, Physics Department, \\
Faculty of Science, Mansoura University, Mansoura 35516, Egypt}
\maketitle

\begin{abstract}
The reductive perturbation method has been employed to derive the
Korteweg--de Vries (KdV) equation for small but finite amplitude
electrostatic ion-acoustic waves in unmagnitized collisionless weakly
relativistic warm plasma. The Lagrangian of the time fractional KdV equation
is used in similar form to the Lagrangian of the regular KdV equation. The
variation of the functional of this Lagrangian leads to the Euler-Lagrange
equation that leads to the time fractional KdV equation. The
Riemann-Liouvulle definition of the fractional derivative is used to
describe the time fractional operator in the fractional KdV equation. The
variational-iteration method given by He is used to solve the derived time
fractional KdV equation. The calculations of the solution with initial
condition $A_{0}\sec $h$(cx)^{2}$ are carried out. The result of the present
investigation may be applicable to some plasma environments, such as
ionosphere.

\begin{description}
\item $\boldsymbol{Keywords:}$ Ion-acoustic waves; Euler-Lagrange equation,
Riemann-Liouvulle fractional derivative, fractional KdV equation, He's
variational-iteration method.

\item \textbf{PACS:} 05.30.Pr, 05.45.-a, 05.45.Yv
\end{description}
\end{abstract}

\section{Introduction}

Nonlinear evolution equations are widely used as models to describe complex
physical phenomena. Because most classical processes observed in the
physical world are nonconservative, it is important to be able to apply the
power of variational methods to such cases. A method used a Lagrangian that
leads to an Euler-Lagrange equation that is, in some sense, equivalent to
the desired equation of motion. Hamilton's equations are derived from the
Lagrangian and are equivalent to the Euler-Lagrange equation. If a
Lagrangian is constructed using noninteger-order derivatives, then the
resulting equation of motion can be nonconservative. It was shown that such
fractional derivatives in the Lagrangian describe nonconservative forces [1,
2]. Further study of the fractional Euler-Lagrange can be found in the work
of Agrawal [3, 4], Baleanu and coworkers [5, 6] and Tarasov and Zaslavsky
[7, 8]. During the last decades, Fractional Calculus has been applied to
almost every field of science, engineering and mathematics. Some of the
areas where Fractional Calculus has been applied include viscoelasticity and
rheology, electrical engineering, electrochemistry, biology, biophysics and
bioengineering, signal and image processing, mechanics, mechatronics,
physics, and control theory [9].

The propagation of solitary waves is important as it describes
characteristic nature of the interaction of the waves and the plasmas. In
the case where the velocity of particles is much smaller than that of light,
ion-acoustic waves present the non-relativistic behaviors, but in the case
where the velocity of particles approaches that of light, the relativistic
effect becomes dominant [10]. Actually high speed and energetic streaming
ions with the energy from 0.1 to 100 $\unit{MeV}$ are frequently observed in
solar atmosphere and interplanetary space. Nevertheless, relativistic
ion-acoustic waves have not been well investigated. When we assume that the
ion energy depends only on the kinetic energy, such plasma ions have to
attain very high velocity of relativistic order. Thus, by considering the
weakly relativistic effect where the ion velocity is about $\frac{1}{10}$\
of the velocity of light, we can describe the relativistic motion of such
ions in the study of nonlinear interaction of the waves and the plasmas
[11]. It appears that the weakly relativistic and ion temperature effects
play an important role to energetic ion-acoustic waves propagating in
interplanetary space [12]. Washimi and Taniti [13] were the first to use
reductive perturbation method to study the propagation of a slow modulation
of a quasimonochromatic waves through plasma. And then the attention has
been focused by many authors [14--15].

To the author's knowledge, the problem of time fractional KdV equation in
weakly relativistic plasma has not been addressed in the literature before.
So, our motive here is to study the effects of time fractional parameter on
the electrostatic structures for a system of collisionless plasma consisting
of a mixture of warm ion-fluid and isothermal electrons with ion flow
velocity has a weakly relativistic effect. We expect that the inclusion of
time fractional parameter and a weakly relativistic effect will change the
properties as well as the regime of existence of solitons.

Several methods have been used to solve fractional differential equations
such as: the Laplace transform method, the Fourier transform method, the
iteration method and the operational method [16]. Recently, there are some
papers deal with the existence and multiplicity of solution of nonlinear
fractional differential equation by the use of techniques of nonlinear
analysis [17-18]. In this paper, the resultant fractional KdV equation will
be solved using a variational-iteration method (VIM) firstly used by He [19].

This paper is organized as follows: Section 2 is devoted to describe the
formulation of the time-fractional KdV (FKdV) equation using the variational
Euler-Lagrange method. In section 3, Variational-Iteration Method (VIM) is
discussed. The resultant time-FKdV equation is solved approximately using
VIM. Section 5 contains the results of calculations and discussion of these
results.

\section{Basic equations and KdV equation}

Consider collisionless ionization-free unmagnetized plasma consisting of a
mixture of warm ions-fluid and isothermal electrons. Assume that the ions
flow velocity has a weak relativistic effect, and therefore there exist
streaming ions in an equilibrium state when sufficiently small- but
finite-amplitude waves propagate one-dimensionally. Such a system is
governed by the following normalized equations [12]:%
\begin{eqnarray}
\frac{\partial }{\partial t}n(x,t)+\frac{\partial }{\partial x}%
[n(x,t)u(x,t)] &=&0\text{,}  \TCItag{1a} \\
\lbrack \frac{\partial }{\partial t}+u(x,t)\frac{\partial }{\partial x}%
][\gamma (x,t)u(x,t)]+\frac{\sigma }{n(x,t)}\frac{\partial }{\partial x}%
p(x,t)+\frac{\partial }{\partial x}\phi (x,t) &=&0\text{,}  \TCItag{1b} \\
\lbrack \frac{\partial }{\partial t}+u(x,t)\frac{\partial }{\partial x}%
]p(x,t)+3p(x,t)\frac{\partial }{\partial x}[\gamma u(x,t)] &=&0\text{,} 
\TCItag{1c} \\
\frac{\partial ^{2}}{\partial x^{2}}\phi (x,t)+n(x,t)-n_{e}(x,t) &=&0\text{.}
\TCItag{1d}
\end{eqnarray}

The electrons temperature $T_{e}$ is much larger than the ions temperature $%
T_{i}$ and in this case, for simplicity, one can neglect the inertia of the
electrons relative to that of the ions, i.e. the high-frequency plasma
oscillations are neglected. Since we are interested in the regime of density
and velocity fluctuations near the ion plasma frequency, so the isothermal
electrons given by:

\begin{equation}
n_{e}(x,t)=\exp [\phi (x,t)]\text{,}  \tag{1e}
\end{equation}%
for weakly relativistic effects, the relativistic factor $\gamma
(x,t)=[1-u(x,t)^{2}/c_{0}^{2}]^{-\frac{1}{2}}$ is approximated by

\begin{equation}
\gamma (x,t)\approx 1+u(x,t)^{2}/(2c_{0}^{2})\text{,}  \tag{1f}
\end{equation}%
where $c$ is the velocity of light. In equations (1) $n(x,t)$\ and $%
n_{e}(x,t)$\ are the densities of ions and electrons, respectively, $u(x,t)$%
\ is the ions flow velocity, $p(x,t)$\ is the ions pressure, $\phi (x,t)$\
is the electric potential, $x$\ is the space co-ordinate and $t$\ is the
time variable. $\sigma =T_{i}/T_{e}<<1$\ is the ratio of the ions
temperature\ to the electrons temperature. All these quantities are
dimensionless, being normalized in terms of the following characteristic
quantities: $n(x,t)$\ and $n_{e}(x,t)$\ are normalized by the unperturbed
electrons density $n_{0}$, $u(x,t)$\ and $c$\ are normalized by the sound
velocity ($k_{B}T_{e}/m_{i})^{1/2},$\ $p(x,t)$ and $\phi (x,t)$\ are
normalized by $n_{0}k_{B}T_{i}$\ and $k_{B}T_{e}/e$, respectively and\ $t$\
and $x$\ \ are normalized by the inverse of the plasma frequency $\omega
_{pi}^{-1}=(4\pi e^{2}n_{0}/m_{i})^{-1/2\text{ }}$and the electron Debye
length $\lambda _{D}=(k_{B}T_{e}/4\pi e^{2}n_{0})^{1/2}$,\ respectively. $%
k_{B}$\ is the Boltzmann's constant, $e$ is the elecron charge and $m_{i}$\
is the mass of plasma ion.

According to the general method of reductive perturbation theory, we
introduce the stretched variables

\begin{equation}
\tau =\epsilon ^{\frac{3}{2}}t\text{,\ \ }\xi =\epsilon ^{\frac{1}{2}%
}(x-\lambda t),  \tag{2}
\end{equation}%
where $\lambda $\ is the the phase velocity. All the physical quantities
appeared in (1) are expanded as power series in terms of the amplitude of
the perturbation $\epsilon $\ about the equilibrium values as%
\begin{eqnarray}
n(\xi ,\tau ) &=&1+\epsilon n_{1}(\xi ,\tau )+\epsilon ^{2}n_{2}(\xi ,\tau
)+...\text{,}  \TCItag{3a} \\
u(\xi ,\tau ) &=&u_{0}+\epsilon u_{1}(\xi ,\tau )+\epsilon ^{2}u_{2}(\xi
,\tau )+...\text{,}  \TCItag{3b} \\
p(\xi ,\tau ) &=&1+\epsilon \text{ }p_{1}(\xi ,\tau )+\epsilon ^{2}p_{2}(\xi
,\tau )+...\text{,}  \TCItag{3c} \\
\phi (\xi ,\tau ) &=&\epsilon \phi _{1}(\xi ,\tau )+\epsilon ^{2}\phi
_{2}(\xi ,\tau )+...\text{,}  \TCItag{3d}
\end{eqnarray}%
with the boundary conditions that as $\left\vert \xi \right\vert \rightarrow
\infty $,\ $n=n_{e}=p=1$, $u=u_{0}$, $\phi =0.\ $

Substituting (2) and (3) into the system of equations (1) and equating
coefficients of like powers of $\epsilon $. Then, from the lowest,
second-order equations in $\epsilon $ and by eliminating the second order
perturbed quantities $n_{2}(\xi ,\tau )$, $u_{2}(\xi ,\tau )$, $p_{2}(\xi
,\tau )$ and $\phi _{2}(\xi ,\tau )$, we obtain the following KdV equation
for the first-order perturbed potential:

\begin{equation}
\frac{\partial }{\partial \tau }\phi _{1}(\xi ,\tau )+A\text{ }\phi _{1}(\xi
,\tau )\frac{\partial }{\partial \xi }\phi _{1}(\xi ,\tau )+B~\frac{\partial
^{3}}{\partial \xi ^{3}}\phi _{1}(\xi ,\tau )=0\text{,}  \tag{4a}
\end{equation}%
where%
\begin{equation}
A=\lambda -\left( \lambda ^{2}-3\sigma \right) \frac{\gamma _{2}}{\gamma _{1}%
}+\frac{3\sigma }{2\lambda }(3\gamma _{1}-1)\text{, \ \ \ \ \ }B=\frac{%
(\lambda ^{2}-3\sigma )}{2\lambda }\text{,}  \tag{4b}
\end{equation}%
with the transendental equation for $\lambda $ as%
\begin{equation}
1-\left( \lambda ^{2}-3\sigma \right) \gamma _{1}=0\text{ \ }\Longrightarrow 
\text{ \ }\lambda =\pm \sqrt{\frac{3\sigma \gamma _{1}+1}{\gamma _{1}}}\text{%
,}  \tag{4c}
\end{equation}%
and%
\begin{equation}
\gamma _{1}=1+\frac{3}{2}R^{2}\text{, \ \ \ \ \ \ \ }\gamma _{2}=\frac{3}{2}%
\frac{R}{c_{0}}\text{, \ \ \ \ \ \ \ }R=\frac{u_{0}}{c_{0}}\text{.}  \tag{4d}
\end{equation}

In equation (4a), $\phi _{1}(\xi ,~\tau )$ is a field variable, $\xi $ is a
space coordinate in the propagation direction of the field and $\tau \in T$ (%
$=[0,T_{0}]$) is the time.f

The resultant KdV equation (4a) can be converted into time-fractional KdV
equation as follows: Using a potential function $V(\xi ,~\tau )$ where $\phi
_{1}(\xi ,~\tau )=V_{\xi }(\xi ,~\tau )=\Phi (\xi ,~\tau )$ gives the
potential equation of the regular KdV equation (1) in the form%
\begin{equation}
V_{\xi \tau }(\xi ,~\tau )+A~V_{\xi }(\xi ,~\tau )v_{\xi \xi }(\xi ,~\tau
)+B~V_{\xi \xi \xi \xi }(\xi ,~\tau )=0\text{,}  \tag{5}
\end{equation}%
where the subscripts denote the partial differentiation of the function with
respect to the parameter. The Lagrangian of this regular KdV equation (4a)
can be defined using the semi-inverse method [20, 21] as follows.

The functional of the potential equation (5) can be represented by%
\begin{equation}
J(V)=\dint\limits_{R}d\xi \dint\limits_{T}d\tau \{V(\xi ,\tau )[c_{1}V_{\xi
\tau }(\xi ,\tau )+c_{2}AV_{\xi }(\xi ,\tau )v_{\xi \xi }(\xi ,\tau
)+c_{3}BV_{\xi \xi \xi \xi }(\xi ,\tau )]\}\text{,}  \tag{6}
\end{equation}%
where $c_{1}$, $c_{2}$ and $c_{3}$ are constants to be determined.
Integrating by parts and taking $V_{\tau }|_{R}=V_{\xi }|_{R}=V_{\xi
}|_{T}=0 $ lead to%
\begin{equation}
J(V)=\dint\limits_{R}d\xi \dint\limits_{T}d\tau \{V(\xi ,\tau )[-c_{1}V_{\xi
}(\xi ,\tau )V_{\tau }(\xi ,\tau )-\frac{1}{2}c_{2}AV_{\xi }^{3}(\xi ,\tau
)+c_{3}BV_{\xi \xi }^{2}(\xi ,\tau )]\}\text{.}  \tag{7}
\end{equation}

The unknown constants $c_{i}~(i=1$, $2$, $3)$ can be determined by taking
the variation of the functional (7) to make it optimal. Taking the variation
of this functional, integrating each term by parts and making the variation
optimum give the following relation%
\begin{equation}
2c_{1}V_{\xi \tau }(\xi ,\tau )+3c_{2}AV_{\xi }(\xi ,\tau )V_{\xi \xi }(\xi
,\tau )+2c_{3}BV_{\xi \xi \xi \xi }(\xi ,\tau )=0\text{.}  \tag{8}
\end{equation}

As this equation must be equal to equation (5), the unknown constants are
given as%
\begin{equation}
c_{1}=1/2\text{, }c_{2}=1/3\text{ and }c_{3}=1/2\text{.}  \tag{9}
\end{equation}

Therefore, the functional given by (7) gives the Lagrangian of the regular
KdV equation as%
\begin{equation}
L(V_{\tau },~V_{\xi },V_{\xi \xi })=-\frac{1}{2}V_{\xi }(\xi ,\tau )V_{\tau
}(\xi ,\tau )-\frac{1}{6}AV_{\xi }^{3}(\xi ,\tau )+\frac{1}{2}BV_{\xi \xi
}^{2}(\xi ,\tau )\text{.}  \tag{10}
\end{equation}

Similar to this form, the Lagrangian of the time-fractional version of the
KdV equation can be written in the form%
\begin{eqnarray}
F(_{0}D_{\tau }^{\alpha }V,~V_{\xi },V_{\xi \xi }) &=&-\frac{1}{2}%
[_{0}D_{\tau }^{\alpha }V(\xi ,\tau )]V_{\xi }(\xi ,\tau )-\frac{1}{6}%
AV_{\xi }^{3}(\xi ,\tau )+\frac{1}{2}BV_{\xi \xi }^{2}(\xi ,\tau )\text{, } 
\notag \\
0 &\leq &\alpha <1\text{,}  \TCItag{11}
\end{eqnarray}%
where the fractional derivative is represented, using the left
Riemann-Liouville fractional derivative definition as [16]%
\begin{eqnarray}
_{a}D_{t}^{\alpha }f(t) &=&\frac{1}{\Gamma (k-\alpha )}\frac{d^{k}}{dt^{k}}%
[\int_{a}^{t}d\tau (t-\tau )^{k-\alpha -1}f(\tau )]\text{, }  \notag \\
k-1 &\leq &\alpha \leq k\text{, }t\in \lbrack a,b]\text{.}  \TCItag{12}
\end{eqnarray}

The functional of the time-FKdV equation can be represented in the form

\begin{equation}
J(V)=\dint\limits_{R}d\xi \dint\limits_{T}d\tau F(_{0}D_{\tau }^{\alpha
}V,~V_{\xi },V_{\xi \xi })\text{,}  \tag{13}
\end{equation}%
where the time-fractional Lagrangian $F(_{0}D_{\tau }^{\alpha }V,~V_{\xi
},V_{\xi \xi })$ is defined by (11).

Following Agrawal's method [3, 4], the variation of functional (13) with
respect to $V(\xi ,\tau )$ leads to%
\begin{equation}
\delta J(V)=\dint\limits_{R}d\xi \dint\limits_{T}d\tau \{\frac{\partial F}{%
\partial _{0}D_{\tau }^{\alpha }V}\delta _{0}D_{\tau }^{\alpha }V+\frac{%
\partial F}{\partial V_{\xi }}\delta V_{\xi }+\frac{\partial F}{\partial
V_{\xi \xi }}\delta V_{\xi \xi }\}\text{.}  \tag{14}
\end{equation}

The formula for fractional integration by parts reads [3, 16]%
\begin{equation}
\int_{a}^{b}dt~f(t)~_{a}D_{t}^{\alpha
}g(t)=\int_{a}^{b}dt~g(t)~_{t}D_{b}^{\alpha }f(t)\text{, \ \ \ }f(t)\text{, }%
g(t)\text{ }\in \lbrack a,~b]\text{,}  \tag{15}
\end{equation}%
where $_{t}D_{b}^{\alpha }$, the right Riemann-Liouville fractional
derivative, is defined by [16]%
\begin{eqnarray}
_{t}D_{b}^{\alpha }f(t) &=&\frac{(-1)^{k}}{\Gamma (k-\alpha )}\frac{d^{k}}{%
dt^{k}}[\int_{t}^{b}d\tau (\tau -t)^{k-\alpha -1}f(\tau )]\text{, }  \notag
\\
k-1 &\leq &\alpha \leq 1\text{, }t\in \lbrack a,b]\text{.}  \TCItag{16}
\end{eqnarray}

Integrating the right-hand side of (14) by parts using formula (15) leads to%
\begin{equation}
\delta J(V)=\dint\limits_{R}d\xi \dint\limits_{T}d\tau \lbrack _{\tau
}D_{T_{0}}^{\alpha }(\frac{\partial F}{\partial _{0}D_{\tau }^{\alpha }V})-%
\frac{\partial }{\partial \xi }(\frac{\partial F}{\partial V_{\xi }})+\frac{%
\partial ^{2}}{\partial \xi ^{2}}(\frac{\partial F}{\partial V_{\xi \xi }}%
)]\delta V\text{,}  \tag{17}
\end{equation}%
where it is assumed that $\delta V|_{T}=\delta V|_{R}=\delta V_{\xi }|_{R}=0$%
.

Optimizing this variation of the functional $J(V)$, i. e; $\delta J(V)=0$,
gives the Euler-Lagrange equation for the time-FKdV equation in the form%
\begin{equation}
_{\tau }D_{T_{0}}^{\alpha }(\frac{\partial F}{\partial _{0}D_{\tau }^{\alpha
}V})-\frac{\partial }{\partial \xi }(\frac{\partial F}{\partial V_{\xi }})+%
\frac{\partial ^{2}}{\partial \xi ^{2}}(\frac{\partial F}{\partial V_{\xi
\xi }})=0\text{.}  \tag{18}
\end{equation}

Substituting the Lagrangian of the time-FKdV equation (11) into this
Euler-Lagrange formula (18) gives%
\begin{equation}
-\frac{1}{2}~_{\tau }D_{T_{0}}^{\alpha }V_{\xi }(\xi ,\tau )+\frac{1}{2}%
~_{0}D_{\tau }^{\alpha }V_{\xi }(\xi ,\tau )+AV_{\xi }(\xi ,\tau )V_{\xi \xi
}(\xi ,\tau )+BV_{\xi \xi \xi \xi }(\xi ,\tau )=0\text{.}  \tag{19}
\end{equation}

Substituting for the potential function, $V_{\xi }(\xi ,\tau )=\phi _{1}(\xi
,\tau )=\Phi (\xi ,\tau )$, gives the time-FKdV equation for the state
function $\Phi (\xi ,\tau )$ in the form%
\begin{equation}
\frac{1}{2}[_{0}D_{\tau }^{\alpha }\Phi (\xi ,\tau )-_{\tau
}D_{T_{0}}^{\alpha }\Phi (\xi ,\tau )]+A~\Phi (\xi ,\tau )~\Phi _{\xi }(\xi
,\tau )+B~\Phi _{\xi \xi \xi }(\xi ,\tau )=0\text{,}  \tag{20}
\end{equation}%
where the fractional derivatives $_{0}D_{\tau }^{\alpha }$ and $_{\tau
}D_{T_{0}}^{\alpha }$ are, respectively the left and right Riemann-Liouville
fractional derivatives and are defined by (12) and (16).

The time-FKdV equation represented in (20) can be rewritten by the formula%
\begin{equation}
\frac{1}{2}~_{0}^{R}D_{\tau }^{\alpha }\Phi (\xi ,\tau )+A~\Phi (\xi ,\tau
)~\Phi _{\xi }(\xi ,\tau )+B~\Phi _{\xi \xi \xi }(\xi ,\tau )=0\text{,} 
\tag{21}
\end{equation}%
where the fractional operator $_{0}^{R}D_{\tau }^{\alpha }$ is called Riesz
fractional derivative and can be represented by [4, 16]%
\begin{eqnarray}
~_{0}^{R}D_{t}^{\alpha }f(t) &=&\frac{1}{2}[_{0}D_{t}^{\alpha
}f(t)+~(-1)^{k}{}_{t}D_{T_{0}}^{\alpha }f(t)]  \notag \\
&=&\frac{1}{2}\frac{1}{\Gamma (k-\alpha )}\frac{d^{k}}{dt^{k}}%
[\int_{a}^{t}d\tau |t-\tau |^{k-\alpha -1}f(\tau )]\text{, }  \notag \\
k-1 &\leq &\alpha \leq 1\text{, }t\in \lbrack a,b]\text{.}  \TCItag{22}
\end{eqnarray}

The nonlinear fractional differential equations have been solved using
different techniques [16-20]. In this paper, a variational-iteration method
(VIM) [21] has been used to solve the time-FKdV equation that formulated
using Euler-Lagrange variational technique.

\section{Variational-iteration method}

Variational-iteration method (VIM) [21] has been used successfully to solve
different types of integer nonlinear differential equations [22, 23]. Also,
VIM is used to solve linear and nonlinear fractional differential equations
[24, 25]. This VIM has been used in this paper to solve the formulated
time-FKdV equation.

A general Lagrange multiplier method is constructed to solve non-linear
problems, which was first proposed to solve problems in quantum mechanics
[21]. The VIM is a modification of this Lagrange multiplier method [22]. The
basic features of the VIM are as follows. The solution of a linear
mathematical problem or the initial (boundary) condition of the nonlinear
problem is used as initial approximation or trail function. A more highly
precise approximation can be obtained using iteration correction functional.
Considering a nonlinear partial differential equation consists of a linear
part $\overset{\symbol{94}}{L}U(x,t)$, nonlinear part $\overset{\symbol{94}}{%
N}U(x,t)$ and a free term $f(x,t)$ represented as%
\begin{equation}
\overset{\symbol{94}}{L}U(x,t)+\overset{\symbol{94}}{N}U(x,t)=f(x,t)\text{,}
\tag{23}
\end{equation}%
where $\overset{\symbol{94}}{L}$ is the linear operator and $\overset{%
\symbol{94}}{N}$ is the nonlinear operator. According to the VIM, the ($n+1$)%
\underline{th} approximation solution of (23) can be given by the iteration
correction functional as [24, 25]%
\begin{equation}
U_{n+1}(x,t)=U_{n}(x,t)+\int_{0}^{t}d\tau \lambda (\tau )[\overset{\symbol{94%
}}{L}U_{n}(x,\tau )+\overset{\symbol{94}}{N}\overset{\symbol{126}}{U}%
_{n}(x,\tau )-f(x,\tau )]\text{, \ \ }n\geq 0\text{,}  \tag{24}
\end{equation}%
where $\lambda (\tau )$ is a Lagrangian multiplier and $\overset{\symbol{126}%
}{U}_{n}(x,\tau )$ is considered as a restricted variation function, i. e; $%
\delta \overset{\symbol{126}}{U}_{n}(x,\tau )=0$. Extreme the variation of
the correction functional (24) leads to the Lagrangian multiplier $\lambda
(\tau )$. The initial iteration can be used as the solution of the linear
part of (23) or the initial value $U(x,0)$. As $n$ tends to infinity, the
iteration leads to the exact solution of (23), i. e;%
\begin{equation}
U(x,t)=\underset{n\rightarrow \infty }{\lim }U_{n}(x,t)\text{.}  \tag{25}
\end{equation}%
\qquad For linear problems, the exact solution can be given using this
method in only one step where its Lagrangian multiplier can be exactly
identified.

\section{Time-fractional KdV equation solution}

The time-FKdV equation represented by (21) can be solved using the VIM by
the iteration correction functional (24) as follows:

Affecting from left by the fractional operator on (21) leads to%
\begin{eqnarray}
\frac{\partial }{\partial \tau }\Phi (\xi ,\tau ) &=&~_{0}^{R}D_{\tau }^{\
\alpha -1}\Phi (\xi ,\tau )|_{\tau =0}\frac{\tau ^{\alpha -2}}{\Gamma
(\alpha -1)}  \notag \\
&&-\ _{0}^{R}D_{\tau }^{\ 1-\alpha }[A~\Phi (\xi ,\tau )\frac{\partial }{%
\partial \xi }\Phi (\xi ,\tau )+B~\frac{\partial ^{3}}{\partial \xi ^{3}}%
\Phi (\xi ,\tau )]\text{, }  \notag \\
0 &\leq &\alpha \leq 1\text{, }\tau \in \lbrack 0,T_{0}]\text{,}  \TCItag{26}
\end{eqnarray}%
where the following fractional derivative property is used [16]%
\begin{eqnarray}
\ _{a}^{R}D_{b}^{\ \alpha }[\ _{a}^{R}D_{b}^{\ \beta }f(t)]
&=&~_{a}^{R}D_{b}^{\ \alpha +\beta }f(t)-\overset{k}{\underset{j=1}{\sum }}\
_{a}^{R}D_{b}^{\ \beta -j}f(t)|_{t=a}~\frac{(t-a)^{-\alpha -j}}{\Gamma
(1-\alpha -j)}\text{, }  \notag \\
k-1 &\leq &\beta <k\text{.}  \TCItag{27}
\end{eqnarray}

As $\alpha <1$, the Riesz fractional derivative $_{0}^{R}D_{\tau }^{\ \alpha
-1}$ is considered as Riesz fractional integral $_{0}^{R}I_{\tau }^{1-\alpha
}$ that is defined by [16]%
\begin{eqnarray}
\ _{0}^{R}I_{\tau }^{\ \alpha }f(t) &=&\frac{1}{2}[_{0}I_{\tau }^{\ \alpha
}f(t)\ +~_{\tau }I_{b}^{\ \alpha }f(t)]  \notag \\
&=&\frac{1}{2}\frac{1}{\Gamma (\alpha )}\int_{a}^{b}d\tau |t-\tau |^{\alpha
-1}f(\tau )\text{, }\alpha >0\text{.}  \TCItag{28}
\end{eqnarray}%
where $_{0}I_{\tau }^{\ \alpha }f(t)$\ and $_{\tau }I_{b}^{\ \alpha }f(t)$
are the left and right Riemann-Liouvulle fractional integrals, respectively
[16].

The iterative correction functional of equation (26) is given as%
\begin{eqnarray}
\Phi _{n+1}(\xi ,\tau ) &=&\Phi _{n}(\xi ,\tau )+\int_{0}^{\tau }d\tau
^{\prime }\lambda (\tau ^{\prime })\{\frac{\partial }{\partial \tau ^{\prime
}}\Phi _{n}(\xi ,\tau ^{\prime })  \notag \\
&&-~_{0}^{R}I_{\tau ^{\prime }}^{1-\alpha }\Phi _{n}(\xi ,\tau ^{\prime
})|_{\tau ^{\prime }=0}\frac{\tau ^{\prime \alpha -2}}{\Gamma (\alpha -1)} 
\notag \\
&&+\ _{0}^{R}D_{\tau ^{\prime }}^{\ 1-\alpha }[A\overset{\symbol{126}}{\Phi
_{n}}(\xi ,\tau ^{\prime })\frac{\partial }{\partial \xi }\overset{\symbol{%
126}}{\Phi _{n}}(\xi ,\tau ^{\prime })+B~\frac{\partial ^{3}}{\partial \xi
^{3}}\overset{\symbol{126}}{\Phi _{n}}(\xi ,\tau ^{\prime })]\}\text{,} 
\TCItag{29}
\end{eqnarray}%
where $n\geq 0$ and the function $\overset{\symbol{126}}{\Phi _{n}}(\xi
,\tau )$ is considered as a restricted variation function, i. e; $\delta 
\overset{\symbol{126}}{\Phi _{n}}(\xi ,\tau )=0$. The extreme of the
variation of (29) using the restricted variation function leads to%
\begin{eqnarray*}
\delta \Phi _{n+1}(\xi ,\tau ) &=&\delta \Phi _{n}(\xi ,\tau
)+\int_{0}^{\tau }d\tau ^{\prime }\lambda (\tau ^{\prime })~\delta \frac{%
\partial }{\partial \tau ^{\prime }}\Phi _{n}(\xi ,\tau ^{\prime }) \\
&=&\delta \Phi _{n}(\xi ,\tau )+\lambda (\tau )~\delta \Phi _{n}(\xi ,\tau
)-\int_{0}^{\tau }d\tau ^{\prime }\frac{\partial }{\partial \tau ^{\prime }}%
\lambda (\tau ^{\prime })~\delta \Phi _{n}(\xi ,\tau ^{\prime })=0\text{.}
\end{eqnarray*}

This relation leads to the stationary conditions $1+\lambda (\tau )=0$ and $%
\frac{\partial }{\partial \tau ^{\prime }}\lambda (\tau ^{\prime })=0$,
which lead to the Lagrangian multiplier as $\lambda (\tau ^{\prime })=-1$. \
Therefore, the correction functional (29) is given by the form%
\begin{eqnarray}
\Phi _{n+1}(\xi ,\tau ) &=&\Phi _{n}(\xi ,\tau )-\int_{0}^{\tau }d\tau
^{\prime }\{\frac{\partial }{\partial \tau ^{\prime }}\Phi _{n}(\xi ,\tau
^{\prime })  \notag \\
&&-~_{0}^{R}I_{\tau ^{\prime }}^{1-\alpha }\Phi _{n}(\xi ,\tau ^{\prime
})|_{\tau ^{\prime }=0}\frac{\tau ^{\prime \alpha -2}}{\Gamma (\alpha -1)} 
\notag \\
&&+\ _{0}^{R}D_{\tau ^{\prime }}^{\ 1-\alpha }[A~\Phi _{n}(\xi ,\tau
^{\prime })\frac{\partial }{\partial \xi }\Phi _{n}(\xi ,\tau ^{\prime })+B~%
\frac{\partial ^{3}}{\partial \xi ^{3}}\Phi _{n}(\xi ,\tau ^{\prime })]\}%
\text{,}  \TCItag{30}
\end{eqnarray}%
where $n\geq 0$.

In Physics, if $\tau $ denotes the time-variable, the right
Riemann-Liouville fractional derivative is interpreted as a future state of
the process. For this reason, the right-derivative is usually neglected in
applications, when the present state of the process does not depend on the
results of the future development [3]. Therefore, the right-derivative is
used equal to zero in the following calculations.

The zero order correction of the solution can be taken as the initial value
of the state variable, which is taken in this case as%
\begin{equation}
\Phi _{0}(\xi ,\tau )=\Phi (\xi ,0)=A_{0}\sec \text{h}(c\xi )^{2}\text{,} 
\tag{31}
\end{equation}%
where $A_{0}=\frac{3v}{A}$ and $c=\frac{1}{2}\sqrt{\frac{v}{B}}$ are
constants and $v$ is the travelling wave velocity.

Substituting this zero order approximation into (30) and using the
definition of the fractional derivative (22) lead to the first order
approximation as%
\begin{eqnarray}
\Phi _{1}(\xi ,\tau ) &=&A_{0}\sec \text{h}(c\xi )^{2}  \notag \\
&&+2A_{0}c~\sinh (c\xi )~\sec \text{h}(c\xi )^{3}  \notag \\
&&\ast \lbrack 4c^{2}B+(A_{0}A-12c^{2}B)\sec \text{h}(c\xi )^{2}]\frac{\tau
^{\alpha }}{\Gamma (\alpha +1)}\text{.}  \TCItag{32}
\end{eqnarray}

Substituting this equation into (30), using the definition (22) and the
Maple package lead to the second order approximation in the form%
\begin{eqnarray}
\Phi _{2}(\xi ,\tau ) &=&A_{0}\sec \text{h}(c\xi )^{2}  \notag \\
&&+2A_{0}c~\sinh (c\xi )~\sec \text{h}(c\xi )^{3}  \notag \\
&&\ast \lbrack 4c^{2}B+(A_{0}A-12c^{2}B)\sec \text{h}(c\xi )^{2}]\frac{\tau
^{\alpha }}{\Gamma (\alpha +1)}  \notag \\
&&+2A_{0}c^{2}\sec \text{h}(c\xi )^{2}  \notag \\
&&\ast \lbrack 32c^{4}B^{2}+16c^{2}B(5A_{0}A-63c^{2}B)\sec \text{h}(c\xi
)^{2}  \notag \\
&&+2(3A_{0}^{2}A^{2}-176A_{0}c^{2}AB+1680c^{4}B^{2})\sec \text{h}(c\xi )^{4}
\notag \\
&&-7(A_{0}^{2}A^{2}-42A_{0}c^{2}AB+360c^{4}B^{2})\sec \text{h}(c\xi )^{6}]%
\frac{\tau ^{2\alpha }}{\Gamma (2\alpha +1)}  \notag \\
&&+4A_{0}^{2}c^{3}\sinh (c\xi )~\sec \text{h}(c\xi )^{5}  \notag \\
&&\ast \lbrack 32c^{4}B^{2}+24c^{2}B(A_{0}A-14c^{2}B)\sec \text{h}(c\xi )^{2}
\notag \\
&&+4(A_{0}^{2}A^{2}-32A_{0}c^{2}AB+240c^{4}B^{2})\sec \text{h}(c\xi )^{4} 
\notag \\
&&-5(A_{0}^{2}A^{2}-24A_{0}c^{2}AB+144c^{4}B^{2})\sec \text{h}(c\xi )^{6}] 
\notag \\
&&\ast \frac{\Gamma (2\alpha +1)}{[\Gamma (\alpha +1)]^{2}}\frac{\tau
^{3\alpha }}{\Gamma (3\alpha +1)}\text{.}  \TCItag{33}
\end{eqnarray}

The higher order approximations can be calculated using the Maple or the
Mathematica package to the appropriate order where the infinite
approximation leads to the exact solution.

\section{Results and discussion}

Numerical studies have been made for a small amplitude ion-acoustic waves in
an unmagnetized collisionless plasma consisting of a mixture of a weak
relativistic warm ion-fluid and isothermal electrons. We have derived the
Korteweg-de Vries equation by using the reductive perturbation method [13].
The time-FKdV equation is derived from the Eular-Lagrangian using Agrawal's
method [4]. The Riemann-Liouvulle fractional derivative is used to describe
the time fractional operator in the FKdV equation. He's
variational-iteration method [21-23] is used to solve the derived time-FKdV
equation.

However, since one of our motivations was to study effects of a relativistic
factor $R=u_{0}/c_{0}$ and time fractional order $\alpha $ on the existence
of solitary waves. The present system admits a solitary wave solution for
any order approximation. In Fig.(1), a profile of the bell-shaped solitary
pulse is obtained. Figure (2) shows that both the amplitude and the width of
the solitary wave increases with the relativistic factor $R$. Also, the
increasing of the time fractional order $\alpha \ $ decreases the soliton
amplitude as shown in Fig (3).

In summery, it has been found that amplitude and width of the ion-acoustic
waves as well as parametric regime where the solitons can exist is sensitive
to the relativistic factor $R$. Moreover, the time fractional order $\alpha $
plays a role of higher order perturbation theory in varying the soliton
amplitude. The application of our model might be particularly interesting in
some plasma environments, such as ionosphere.\bigskip \pagebreak

\textbf{Figure Captions}

\textbf{Fig (1):} The electrostatic potential, $\Phi (\xi ,\tau )$, vs $\xi $
and $\tau $, for $v=0.04$, $\sigma =0.03$, $\alpha =0.2$ and $R=0.03$.

\textbf{Fig (2):} The electrostatic potential, $\Phi (\xi ,\tau )$, vs $\xi $
at $\tau =5$, $v=0.04$, $\sigma =0.03$ and $\alpha =0.2$ for different
values of $R$.

\textbf{Fig (3):} The amplitude of electrostatic potential, $\Phi (0,\tau )$%
, vs $\alpha $ at $v=0.04$, $\sigma =0.03$ and $R=0.03$ for different values
of $\tau $.\pagebreak


\begin{thebibliography}{99}
\bibitem{} Riewe, F., Nonconservative Lagrangian and Hamiltonian mechanics,
Physical Review E 53 (1996) 1890.

\bibitem{} Riewe, F., Mechanics with fractional derivatives, Physical Review
E 55 (1997) 3581.

\bibitem{} Agrawal, O. P., Formulation of Euler-Lagrange equations for
fractional variational problems, J. Mathematical Analysis and Applications
272 (2002) 368.

\bibitem{} Agrawal, O. P., Fractional variational calculus in terms of Riesz
fractional derivatives J. Physics A: Mathematical and Theoretical 40 (2007)
6287.

\bibitem{} Baleanu, D. and Avkar, T., Lagrangians with linear velocities
within Riemann-Liouville fractional derivatives, Nuovo Cimento B 119 (2004)
73-79.

\bibitem{} Muslih, S. I., Baleanu, D. and Rabei, E., Hamiltonian formulation
of classical fields within Riemann-Liouville fractional derivatives, Physica
Scripta 73 (2006) 436-438.

\bibitem{} Tarasov, V. E. and Zaslavsky, G. M., Fractional Ginzburg-Landau
equation for fractal media, Physica A: Statistical Mechanics and Its
Applications 354 (2005) 249-261.

\bibitem{} Tarasov, V. E. and Zaslavsky, G. M., Nonholonomic constraints
with fractional derivatives, J. Physics A: Mathematical and General 39
(2006) 9797-9815.

\bibitem{} Sabatier, J., Agrawal, O. P. and Tenreiro Machado, J. A.
(editors), Advances in Fractional Calculus, (Springer, Dordrecht, The
Netherlands, 2007).

\bibitem{} Das, G. C. and Paul, S. N., Phys. Fluids 28 (1985) 823.

\bibitem{} Nejoh, Y., The effect of the ion temperature on the ion acoustic
solitary waves in a collisionless relativistic plasma, J. Plasma Physics
37(3) (1987) 487-495.

\bibitem{} El-Labany, S. K., Contribution of higher-order nonlinearity to
nonlinear ion-acoustic waves in a weakly relativistic warm plasma. Part 1.
Isothermal case, J. Plasma Physics 50 (1993) 495.

\bibitem{} Washimi, H. and Taniuti, T., Propagation of Ion-Acoustic Solitary
Waves of Small Amplitude, Physical Review Letters 17 (1966) 996-998.

\bibitem{} Taniuti, T. and Wei, C-C., Reductive Perturbation Method in
Nonlinear Wave Propagation. I, J. Physical Society Japan 24 (1968) 941-946.

\bibitem{} El-Shewy, E. K., Zahran, M. A., Schoepf, K., Elwakil, S. A.,
Contribution of higher order dispersion to nonlinear dust-acoustic solitary
waves in dusty plasma with different sized dust grains and nonthermal ions,
Physica Scripta 78 (2008) 025501.

\bibitem{} Podlubny, I., Fractional Differential Equations, (Academic Press,
San Diego, 1999).

\bibitem{} Babakhani, A. and Gejji, V. D., Existence of positive solutions
of nonlinear fractional differential equations, J. Mathematical Analysis and
Applications 278 (2003) 434--442.

\bibitem{} He, J.-H., Approximate analytical solution for seepage flow with
fractional derivatives in porous media, Computer Methods in Applied
Mechanics and Engineering 167 (1998) 57-68.

\bibitem{} He, J.-H., A new approach to nonlinear partial differential
equations, Communication Nonlinear Science and Numerical Simulation 2(4)
(1997) 230-235.

\bibitem{} He, J.-H., Semi-inverse method of establishing generalized
variational principles for fluid mechanics with emphasis on turbo-machinery
aerodynamics, Int. J. Turbo Jet-Engines 14(1) (1997) 23-28.

\bibitem{} He, J.-H., Variational principles foe some nonlinear partial
differential equations with variable coefficients, Chaos, Solitons and
Fractals 19 (2004) 847-851.

\bibitem{} He, J.-H. and Wu, X.-H., Construction of solitary solution and
compacton-like solution by variational iteration method, Chaos, Solitons and
Fractals 29 (2006) 108-113.

\bibitem{} Abulwafa, E. M., Abdou M. A., Mahmoud A. A., The
Variational-Iteration Method to Solve the Nonlinear Boltzmann Equation,
Zeitschrift f\"{u}r Naturforschung A 63a (2008) 131-139.

\bibitem{} Molliq R, Y., Noorani, M. S. M. and Hashim, I., Variational
iteration method for fractional heat- and wave-like equations, Nonlinear
Analysis: Real World Applications 10 (2009) 1854-1869.

\bibitem{} Sweilam, N. H., Khader, M. M. and Al-Bar, R. F., Numerical
studies for a multi-order fractional differential equation, Physics Letters
A 371 (2007) 26--33.\pagebreak
\end{thebibliography}
\end{document}